\documentclass[journal=jacsat,manuscript=article]{achemso}

\usepackage[version=3]{mhchem} 

\author{Kirby Schmidt}
\author{Anthony Trofe}
\author{Tetyana Ignatova}
\email{t_ignato@uncg.edu}
\affiliation{Department of Nanoscience, Joint School of Nanoscience and Nanoengineering,University of North Carolina at Greensboro, Greensboro, 27401, United States}
\phone{+123 (0)123 4445556}

\title{Multimodal Image Registration of Raman Spectral Maps in Two Dimensional Materials by Strain and Doping Analysis}


\begin{document}

\begin{abstract}
It is common to measure a single sample using multiple different microscopy methods that have variable scales, rotation and translation. Registering hyperspectral images of two dimensional materials is particularly difficult due to the lack of keypoints on unprepared substrates. Identifying variations in the strain of these samples can assist in the registration of these samples by creating keypoints to correlate images. Registration of these images allow for multimodal analysis from these various instruments by aligning multiple images into a single coordinate space. This is done by Hough transformations and arbitrary resolution definitions to generate a new coordinate frame where spatial information may be preserved and correlated on a pixel by pixel basis. Such multimodal image alignment may be used to correlate data from various instruments. Strain information is extracted from the Raman spectra and the resulting hyperspectral image is used to register the Raman information with the other modes of microscopy.
\end{abstract}


\section*{Introduction}
Since the discovery of graphene the preparation and classification of two-dimensional (2D) materials has seen large improvements\cite{Geim2007,Bao2011,Kim2017,Gao2020,Wang2021,Lemme2022,Hu2022}. Analysis of these materials is done on a variety of instruments that output information at different resolutions, physical scale and orientation. Correlating information between images from different instruments is a labor intensive process that involves overlaying images and is prone to error and guess work. The issue increases when different methods do not closely resemble one another as is the case with Raman vs. electron microscopy characterization. In this work we find similarities between microscopic images and Raman spectral maps by extracting strain and doping information from the Raman spectra to which reveals similar structures that are otherwise not visible.

Often classification of 2D samples must be done manually by locating and characterizing these materials via Raman spectroscopy, atomic force microscopy (AFM) and scanning electron microscopy (SEM) to determine layer count and quality of these materials. This is a time intensive process that requires trained personnel. Subsequent processing includes projection of Raman maps onto AFM or SEM images to correlate data, requiring further manual alignment. This process is a good candidate to be improved by machine vision technology. Due to diffraction limited resolution Raman spectral maps are at a much lower spacial resolution and often do not show the same patterns as AFM/SEM images. We found that by analyzing the strain and doping of 2D materials these characteristic features appear in the Raman spectral images. The techniques described in this paper will allow one to register multimodal images containing Raman spectral maps allowing for accurate coanalysis and simplifying the task of building large libraries for machine learning approaches in the future.

Raman spectroscopy registration of monolayered materials can be particularly difficult since these materials are often very uniform optically and microscopy images (Figure 1 A,B) do not share features with the Raman spectral maps (Figure 1 C-E,H-J). While some features correspond to the wrinkles in the SEM image and AFM topography scan, no one image has enough information to register the Raman map with either SEM or AFM image. Extracting strain and doping information from the Raman spectral data (Figure 1 F,G) allows us to more accurately register images by revealing features that are not present in other characteristic measures. Defected areas in the graphene like wrinkles and cracks often show a large shift in the strain or doping of the materials, and these wrinkles are visible in the microscopy images. The regions with large strain or doping variations appears with high contrast in the Raman spectral maps. Much research of hyperspectral image registration is done in the medical field where high resolution Raman maps of tissue samples can be overlayed with high resolution MRI or optical data. There are two primary ways to register images, feature based methods and intensity based methods.  Feature based methods extract some information from the image, such as key points, edges, or corners. Researchers may implant markers in their tissue samples before beginning, giving key points to reference \cite{Mu_Fu_Kuduvalli_2008, Orchard} or they may use automatic identification of key points by using the SIFT or SURF algorithms \cite{Cheung_Hamarneh_2007, Yu_Yang_Yang_Leng_Cao_Wang_Tian_2016}. These detect key points such as edges and corners and register the key points between images. These methods require distinct features for the best results. Intensity based methods are most often used when key points are difficult to determine. They attempt to match pixel intensities by minimizing some metric like mutual information \cite{Klein_Gigler_Aschenbrenner_Monetti_Bunk_Jamitzky_Morfill_Stark_Schlegel_2012} or cross correlation \cite{Avants_Epstein_Grossman_Gee_2008}.

In this work, data from different instrumentation are set into a similar coordinate base and resolution matched using ridge detection and Hough transformations. This study provides an excellent method for the coanalysis of Raman spectroscopy mapping, atomic force microscopy and other microscopic techniques using strain and doping analysis to register Raman spectroscopic map as key points.

\begin{figure}[ht b]
    \centering
    \includegraphics[width=16cm]{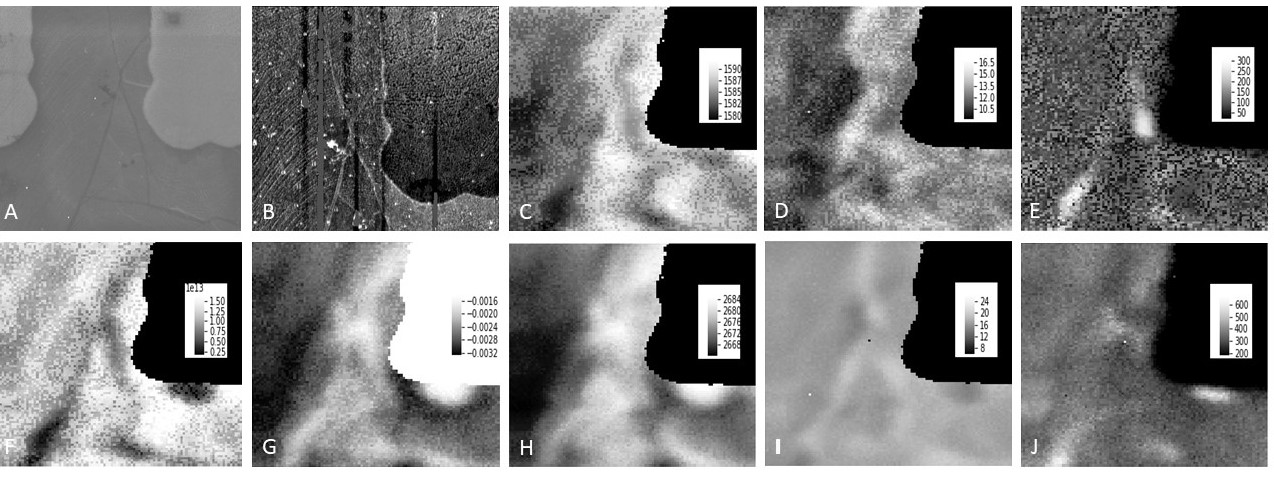}
    \caption{A) SEM image of GFET sample. B) \( 10\times10\: \mu m^2 \) AFM image of GFET sample. C-E) graphene G Raman peak center, width and amplitude respectively. F)  graphene doping extracted from Raman Map. G) graphene strain extracted from Raman map of sample. H-J) graphene 2D Raman peak center, width and amplitude respectively. It is clear from these images that the characteristic G and 2d peaks of graphene do not give enough information to register all images. However, strain and doping maps provide enough information to register all images.}
    \label{fig:1}
\end{figure}

\section{Methods}
\subsection{Strain and Doping Analysis}
Strain and doping information can be extracted from Raman spectra for a number of materials including graphene \cite{Mueller}, \ce{MoS2} \cite{Velicky}, \ce{WSe2} \cite{Kolesnichenko_Zhang_Yun_Zheng_Fuhrer_Davis_2020} as well as other transition metal dichalcogenides \cite{schmidt,Iqbal_Shahzad_Akbar_Hussain_2020}. The process is similar for all of these materials, and follows the process put forth by Mueller et al \cite{Mueller}. In graphene, strain and doping will shift the 2D and G Raman peaks linearly with increasing strain and doping. The shift of both peaks can be characterized by the ratio \(\Delta \omega^h_{2d} / \Delta \omega_g^h \)and \(\Delta \omega^b_{2d} / \Delta \omega_b^h \) where \(\omega\) is the peak center, h is the influence due to hydrostatic strain, and d is the influence due to doping. The vector \(\Vec{O} = (\omega_{2d}^0 / \omega_{g}^0)\) represents the peak location of an unstrained and undoped spectra of graphene. The shift away from \(\Vec{0}\) can be represented by a linear combination of \(\Delta \omega^h_{2d}/ \Delta \omega_g^h \)and \(\Delta \omega^h_{2d}/ \Delta \omega_g^h \) where \(\omega\), whose magnitudes correspond to the magnitude of strain and doping respectively. 

\subsection{Image Preparation}
First, Raman hyperspectral maps must be converted to images that will undergo image registration. The wrinkles in 2D  material create doping effects that radiates out from the wrinkle. We can detect where the wrinkle is by applying a ridge filter. This will calculate the center of the wrinkle using the eigenvalues of the Hessian matrix. Then binarize the image and remove noise by masking over areas of low strain. 
Images, which are all gray scale at this point, are processed using median or bilateral filtering based on image quality, resolution and contrast. For samples with good contrast and high resolution, median filtering can be applied. This technique reduces shot-noise by replacing the center pixel of the filtering kernel by the median pixel in the kernel. Due to this however, ridges can be moved a few pixels so this technique should only be used on high resolution images. If contrast or resolution is lower, bilateral filtering should be applied. Bilateral filtering applies a Gaussian filter over largely uniform areas, while preserving edges. This filtering will also reduce noise, but leave the edge positions intact. We use median filtering for AFM and SEM images, while optical images are filtered using the bilateral filter. Raman-spectra mapped images are not filtered due to their low resolution. 

Binary images must be created after the filtering. Each image has a ridge filter applied by analyzing the Hessian matrix. The maxima of this matrix represents the center of ridges. Thresholding is then applied, creating a binary image where white pixels represent the wrinkles in the graphene. A series of opening and closing morphological operations can be used to smooth the edges. 

A Hough transformation is applied to the binarized images as described by (Hough Transform Paper) to transform our ridge pixel coordinates into parameter space of $x \sin(\theta) + y \cos(\theta) = R (\theta)$. The pixel coordinates are transformed to sinusoidal functions and plotted on a 2D histogram. The largest bins are along the intersections of these functions. The location of the largest bin describes the equation of a line corresponding to a ridge. By selecting a number of the largest bins we extract the longest strait wrinkles in the pictures. These maxima may be used as key points for our transformation. The selection of these key points is critical to successfully aligning images. Wrinkles must be long enough to appear in the Hough transformation, and should be straight. 

\subsection{Registration}
In order to register images extrinsic or intrinsic information is required. Extrinsic information requires the use of markings placed on the substrate which adds time, complexity and cost to a project. Intrinsic information would take the form of information inside of the image which we can use to link images such as key points from edges and corners, or intensity from hyperspectral information. Mutual Information is a intensity metric of two images which is maximized to register two images. This is often the preferred method of image registration in most applications where the images are of similar scale, and the images may contain similar information. However because these hyperspectral images are often very different in resolution and information, Mutual Information maximization is not a good candidate for registration. Raman spectral maps of heterostructures at the small scale often have large intensity shifts over micron distances which do not appear in microscopy images. This makes intensity based matching problematic.  Key point identification is the method we wish to pursue. 

Because 2D materials are nearly uniform over their surface it can be difficult to register two images. If a given sample was perfectly uniform over the entire image area, it would be impossible to register. However, most two dimensional materials are non-uniform at the smallest scales where we can detect wrinkles, tears and defects in the materials. These small defects show brightly in the hyperspectral Raman images after strain and doping is extracted. Two dimensional materials that are folded or wrinkled show large strain or doping variations around these defects, and these defects can be used for image registration. We can find the maxima of this strain variation in the Raman spectral map and use these as 'edges' for keypoint matching.  Wrinkles in an SEM and AFM image align well with the strain maxima. We can transform the spacial coordinates of the wrinkles and strain into Hough space. The units of Hough space are r and theta, and every point in Hough space describes a line in real space with the equation $x \sin(\theta) + y \cos(\theta) = R (\theta)$. A two dimensional histogram is generated by the sinusoids in Hough space. The largest bin in this histogram describes a line in real space. Because of our preprocessing this line will match strait sections in the wrinkles of our two dimensional materials.

\begin{figure}[ht]
    \centering
    \includegraphics[width=12cm]{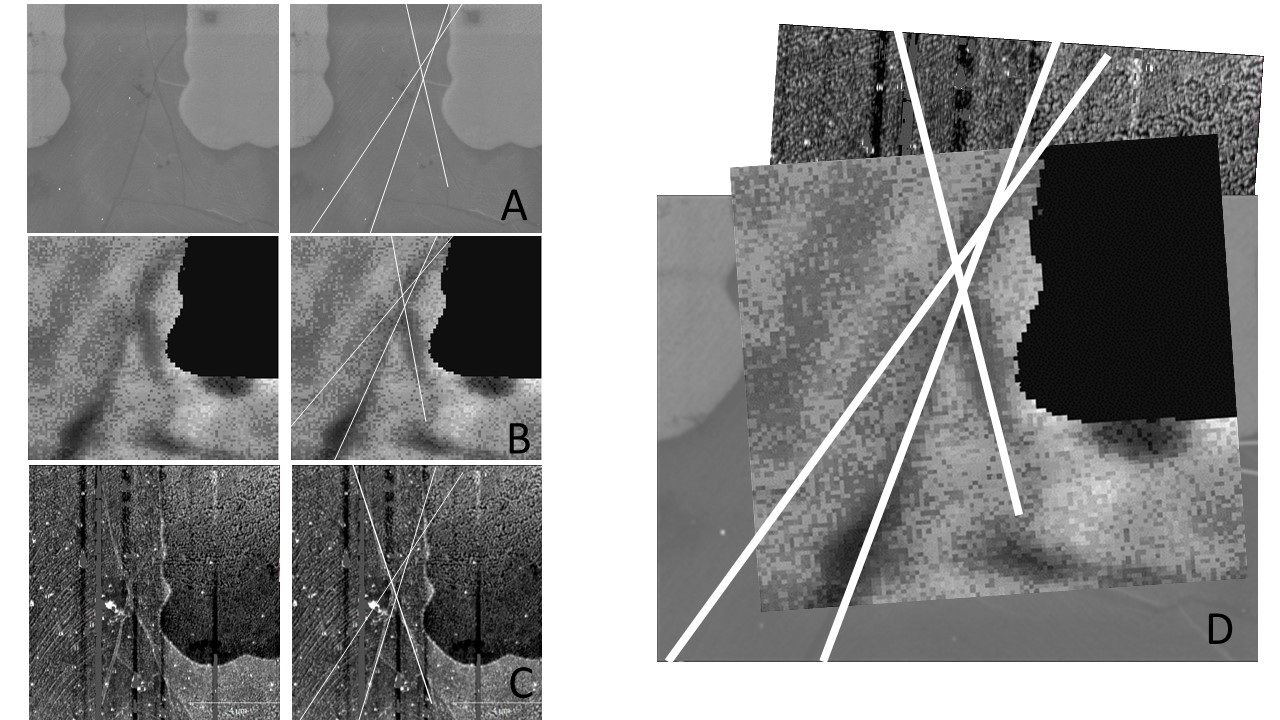}
    \caption{A) SEM image with detected wrinkles on right. B) Raman image, with detected lines on right C) AFM image, with detected lines on right. D) Aligned images by Hough transform.}
    \label{fig:2}
\end{figure}

\subsection{Alignment}
Using the spacial coordinates of wrinkles, we can determine rotation of the images by following the procedure presented by Chitsobhuk et al\cite{Chitsobhuk}. Rotation of an image in Hough space is represented by a shift along the X axis. By aligning our key points vertically we determine the rotation of the image. We can also use phase correlation to determine the translation. By transforming the spacial coordinates by a discrete Fourier transform we determine translation by:

\begin{equation}
    f_2(x,y) = f_1(x-t_x,y-t_y)
\end{equation}

\begin{equation}
    F_2(\psi, \nu) = e^{i2\pi(\psi t_x + \nu t_y)} F_1(\psi, \nu)
\end{equation}

where \(f_1\) and \(f_2\) are signals related by a translation \((t_x\) and \(t_y)\), \((F_1\) and \(F_2\) are their Fourier transformations, the transformations will have the same intensity but will be shifted by the phase \(e^{i2\pi(\psi t_x + \nu t_y)}\)

\subsection{Sample Preparation}
The graphene field-effect transistor (GFET) used was the GFET-S10 acquired from Graphenea. This device hosts 36 individual graphene devices patterend in a grid, 30 of which have Hall-bar geometry and six a 2-probe geometry. Additionally, the dimensions of the graphene channels vary allowing for a range of measurements of property dependency. Specifications include a oxide thickness of 90 nm, Surface dielectric constant of 3.9, and surface resistivity ranging from 1-10 $\Omega$. SEM paramaters for image acquisition of GFET include an accelerating voltage of 3.0 kV and a working depth of 3.4mm.

The equipment used in the experiment include Horiba Raman Confocal Microscope with 532 nm Laser excitation, Zeiss Auriga FIBFESEM, and Asylum MFP-3D Origin+ AFM. This method requires two criteria to be met in order to perform alignment. Crystal edges must be distinguishable and images must be roughly the same size and scale. Raman features must be fit and converted into an image prior to alignment.

Samples analyzed with Raman Spectroscopy are fit using Lorentzian Peaks using the non-linear least-squares minimization curve-fitting python library (LMFIT). For graphene flakes these include D, G, and 2D peaks at $\sim$1350 cm$^{-1}$, $\sim$1580 cm$^{-1}$ and $\sim$2690 cm$^{-1}$ respectively. 

\section*{Summary}
Because of the lack of feature correlation between microscopy methods and Raman spectral mapping it can be difficult to register Raman maps of two dimensional materials. With this method, a Raman map of graphene has been registered with SEM and AFM images after extracting strain and doping information from the Raman spectroscopic map. This technique can also be applied to other two dimensional materials where strain and doping information can be extracted, such as \ce{MoS2} \cite{Velicky}, \ce{WS2} \cite{Kolesnichenko_Zhang_Yun_Zheng_Fuhrer_Davis_2020}, and other transition metal dichalcogenides \cite{Iqbal_Shahzad_Akbar_Hussain_2020}. 

\section*{Acknowledgements}

This work was performed at the Joint School of Nanoscience and Nanoengineering (JSNN), a member of the Southeastern Nanotechnology Infrastructure Corridor (SENIC) and National Nanotechnology Coordinated Infrastructure (NNCI), which is supported by the National Science Foundation (Grant ECCS-1542174). 

\bibliography{sample}

\end{document}